\documentclass[iop,apj]{emulateapj}
\usepackage{amsmath,amssymb,amstext}

\usepackage[breaklinks,colorlinks,citecolor=blue,linkcolor=magenta]{hyperref} 

\usepackage[all]{hypcap} 

\usepackage{natbib}
\usepackage{graphicx}
\usepackage[caption=false]{subfig}
\usepackage{booktabs}

\shorttitle{Short Title}
\shortauthors{Krief et al.}

\begin{document}
	
\title{Line broadening and the solar opacity problem}

	
	\author{M. Krief, A. Feigel and D. Gazit}
	\email{menahem.krief@mail.huji.ac.il}
	\affil{The Racah Institute of Physics, The Hebrew University, 91904 Jerusalem, Israel}
	
\begin{abstract}
	The calculation of line widths constitutes theoretical and computational challenges in the calculation of opacities of hot dense plasmas. Opacity models use line broadening approximations that are untested at stellar interior conditions.
	Moreover, calculations of atomic spectra  of the sun, indicate a large discrepancy in the K-shell line widths between several atomic codes and the OP.
	 	In this work, the atomic code STAR is used to study the sensitivity of solar opacities to line-broadening.
	Variations in the solar opacity profile,
	due to  an increase of the Stark widths resulting from discrepancies with OP, are compared, in light of the solar opacity problem, with the required opacity variations of the present day sun, as imposed by helioseismic and neutrino observations.	 
	 The resulting
	  variation profile, is much larger than the discrepancy between different atomic codes, agrees qualitatively with the missing opacity profile, recovers about half of the missing opacity nearby the convection boundary and has a little effect in the internal regions. Since it is hard to estimate quantitatively the uncertainty in the Stark widths, we show that an increase of all line widths by a factor of about $ \sim 100 $ recovers quantitatively the missing opacity.
	  These results emphasize the possibility that photoexcitation processes are not modeled properly, and more specifically,
	  highlight the need for a better theoretical characterization of the line broadening phenomena at stellar interior conditions and of the uncertainty due to the way it is implemented by atomic codes.
	 	\end{abstract}



\maketitle
	
\section{Introduction}
The revision of the solar photospheric abundances over the past decade
(\cite{asplund2005solar,asplund2009_nonarxiv,caffau2011solar,scott2015,scott2015_2,grevesse2015}),
gave rise to a new problem in solar physics.
The indicated solar metallicity, which is mainly due to low-Z metallic elements, has been significantly revised
downward than previously assumed (\cite{grevesse1993cosmic,grevesse1998standard}).
Using these revised abundances,
standard solar models (SSMs), which are the fundamental theoretical tools to investigate the properties of the solar interior, are in conflict with  helioseismic measurements (\cite{basu2004constraining}), such as the radius of the convection-zone boundary (CZB), the surface helium abundance and the sound speed profile (\cite{serenelli2009new}). The resulting \textit{solar composition problem}, has triggered a rapid increase of research efforts in the field.

The radiative opacity of the solar mixture is a key quantity describing the coupling between radiation and matter in the hot and dense solar interior. 
Absorption by photoexcitation and photoionization can become a major source of opacity for the partially ionized metals,
which contribute significantly to the opacity, although they are only present as a few percent of the mixture. As a result, the opacity profile of the sun depends strongly on the metallic abundances,
and there exists a direct relation between the solar composition problem and the role of metallic opacity in solar models.
Several explanations to the solar composition problem were proposed
(see \cite{Bergemann2014} and references therein).
One popular explanation suggests that the solar opacity profile must be revised.
Indeed, it has been shown (\cite{christensen2009opacity}) that in order to reproduce the helioseismic measurements, changes in the solar opacity profile are required to compensate the revised lower solar metallicity.
It is believed that the opacities of metals in the stellar mixture should be revised upward to compensate for the decreased low-Z metallic element abundances. 
The effect of intrinsic opacity changes (that is, opacity changes with a fixed composition) on observable quantities, i.e. the convection boundary radius, surface helium abundance, sound speed profile and neutrino fluxes, can be studied by the linear solar model (LSM) of \cite{villante2010linear}. 
It was shown
(\cite{serenelli2009new,villante2010constraints,villante2010linear,villante2014chemical,villante2015quantitative}), that a smooth increase of
the opacity from the range of $ 5\%-10\% $ in the central regions to
the range of $ 20\%-30\% $ near the CZB
is required reproduce the helioseismic and neutrino fluxes observations.

Over the past decades, detailed calculations of solar opacities were performed by several groups with state of the art atomic codes, such as the Opacity-Project (OP) (\cite{seaton1994opacities,badnell2005updated,delahaye2016ipopv2}), OPAL (\cite{rogers1992radiative,iglesias1996updated,iglesias2015iron}), ATOMIC (\cite{neuforge2001helioseismic,colgan2013light,fontes2015alamos,colgan2016new}) and OPAS (\cite{blancard2012solar,mondet2015opacity,le2015first}). We have recently used our 
opacity code STAR,  
which implements the Super-Transition-Array (STA) method (\cite{BarShalom1989}),
for a detailed calculation of the solar opacity (\cite{krief2016sun_sta_apj}).

The calculation of spectral line shapes constitutes a theoretical as well as a computational challenge in the calculation of atomic spectra.
The broadening of spectral lines in hot dense plasmas
is due to radiation damping, thermal
Doppler effects, and pressure (Stark) broadening.
The latter stems from complex interactions of the emitting ions with the plasma environment.
These complex interactions are not taken into account in the atomic calculation (i.e. oscillator strength, transition energies, etc.), and affect the spectral opacity mainly through the line-shapes. 
It is well known that even state of the art computer-simulation Stark width models disagree by at least a factor of 2 (\cite{stambulchik2013review,alexiou2014second}).
There is also a disagreement between advanced convergent-close-coupling quantum mechanical line-width calculations and experiments, even for simple Li and Be-like ions for $ Z\leq10 $. It appears that this disagreement increases with the atomic number $ Z $, and, as was shown by \cite{ralchenko2003electron},
the experimental line widths are always
larger than the quantum-mechanical ones, suggesting some additional line broadening mechanism for higher-$ Z $. 
In any case, the experimental data is  only available for much lower temperatures and densities than those found in the solar interior.
In addition, line-wing models, which may have a major effect on Rosseland opacities, are untested experimentally.
Excited spectator electrons also pose computational complications
(\cite{iglesias2010excited,iglesias2015iron}) and
the effect of autoionizing states is still not well understood (\cite{nahar2011highly}).
Moreover, since calculations of spectral opacities for mid and high $ Z $ elements typically include a huge number of spectral lines, simplified line-profiles and approximate line widths are used in opacity codes, in order to reduce the large computational complexity.
More specifically, opacity codes often use Voigt (or modified Voigt \cite{iglesias2009frequency}) profiles, while 
empirical formulae, which provide fits to available experimental data 
(\cite{dimitrijevic1987simple}) or ab-initio calculations 
(\cite{seaton1987atomic}), are used in the calculation of line widths, based on a theory originally proposed by \cite{griem1968semiempirical}. For one, two, and three electron ions, several codes use an approximate linear Stark model by \cite{lee1988plasma} (see also \cite{iglesias2016comparison}) for the calculation of the line profile.

Hence, all opacity models use line broadening approximations that are untested at stellar interior conditions.
Sensitivity studies of solar opacities with respect to the line  widths were performed in the past by
\cite{rogers1992radiative} and also mentioned by \cite{seaton1994opacities,fontes2015relativistic,mondet2015opacity,iglesias2015iron}.
In this work, we use the atomic code STAR to analyze the sensitivity of solar opacities to Stark widths. We use a SSM calibrated with the recent AGSS09 photospheric abundances.
We show that variations in the solar opacity profile due to changes in the Stark widths are mainly due to the K-shell lines.
We compare our results with the constraints on the opacity profile of the present day sun, as imposed in the framework of the linear solar model.

\section{Calculations and results}
\begin{figure}[h]
	\centering
	\resizebox{0.5\textwidth}{!}{\includegraphics{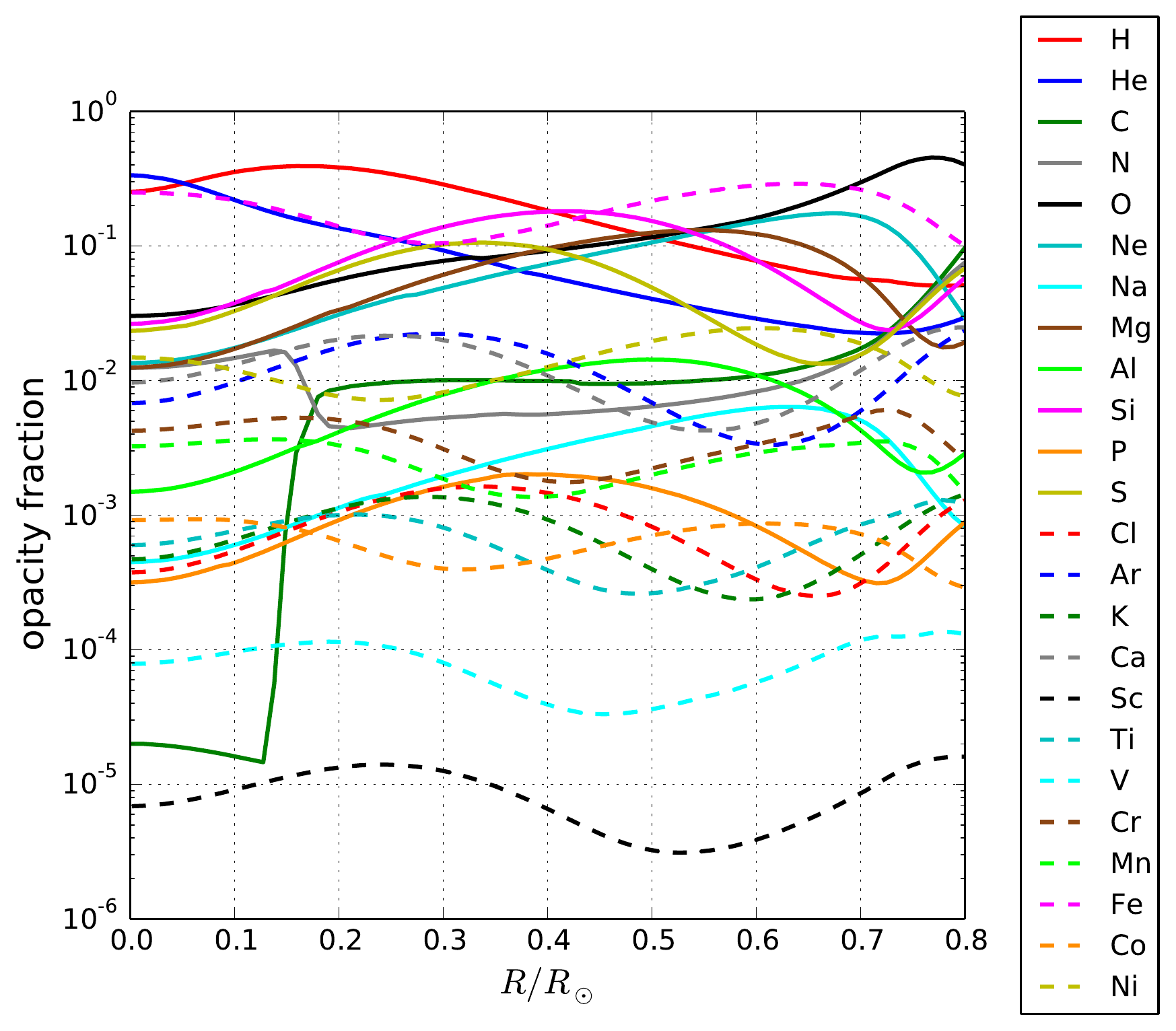}}
	\resizebox{0.5\textwidth}{!}{\includegraphics{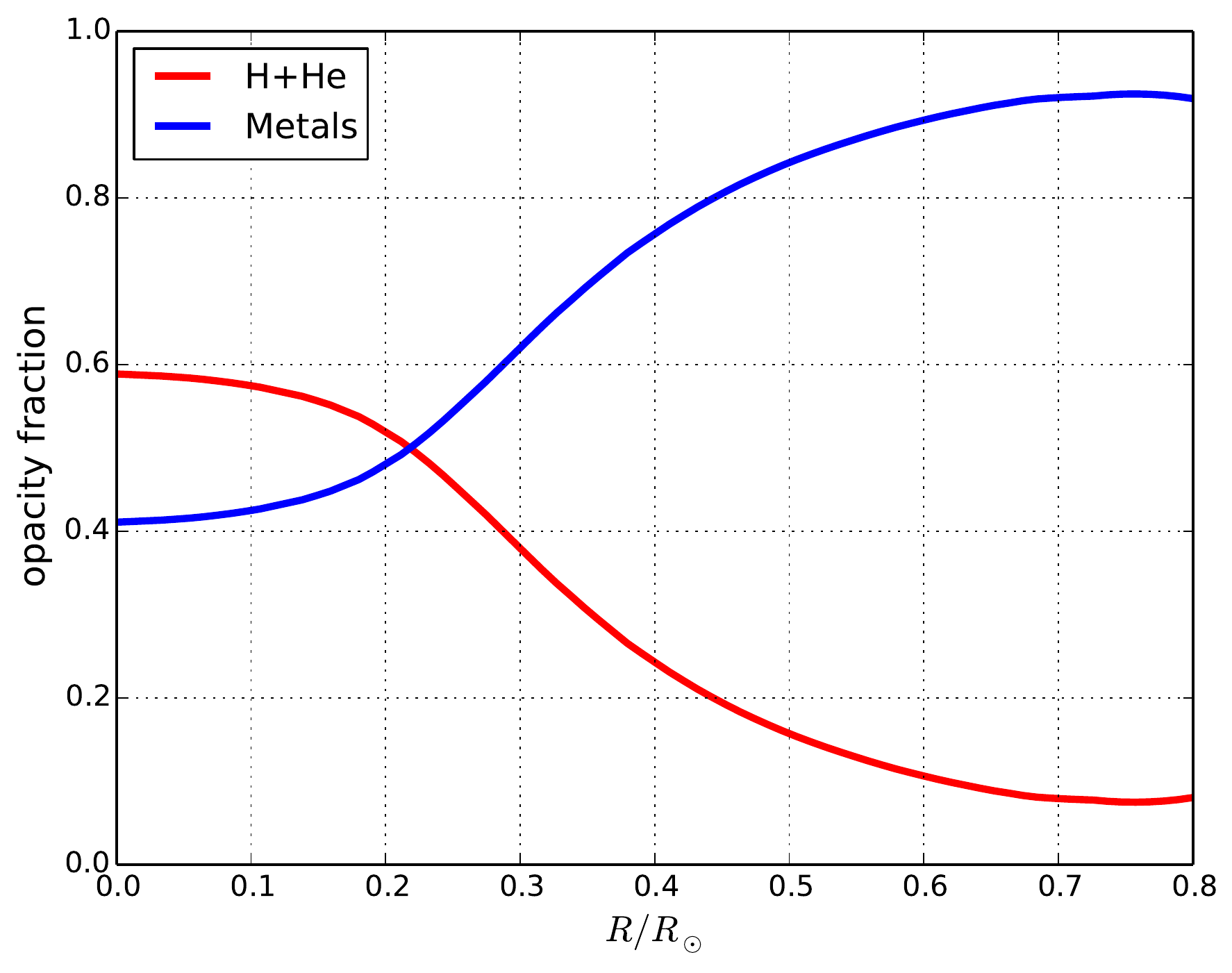}}
	\caption{Rosseland opacity fractions across the solar interior, calculated by STAR. Upper panel: individual contributions of the 24 elements in the solar mixture. Lower panel: the opacity fraction of hydrogen and helium and that of the metals (that is, all the other elements in the mixture).
	Rosseland opacity fractions are calculated as explained in \cite{krief2016sun_sta_apj}.
		}
	\label{fig:elements_contri_kappa}
\end{figure}

The atomic code STAR (\cite{krief2015effect,krief2016sun_sta_apj})
is used 
to examine the sensitivity of the solar opacity profile, to the values of the Stark widths.
Photoexcitation and photoionization atomic calculations are based on a fully relativistic quantum mechanical theory via the Dirac equation.
The inverse bremsstrahlung is calculated via a screened hydrogenic approximation with a degeneracy correction and the Thomson scattering includes corrections for (relativistic) degeneracy, collective and finite-temperature relativistic effects.
The calculations are carried along a thermodynamic path obtained from a solar model
implemented by
\cite{villante2014chemical} (and references therein, see also \cite{krief2016sun_sta_apj}). This solar model is calibrated with the recent 
AGSS09 
set of chemical abundances (\cite{asplund2009_nonarxiv}).
We note that even though the opacity used in this solar model was calculated with the OP atomic code (\cite{seaton1994opacities,badnell2005updated}),
calculation of opacities with different opacity models on the same thermodynamic path is justified in the framework of the linear-solar-model and yields the intrinsic opacity changes of the nominal solar model
(\cite{villante2010linear}).
The solar mixture in the solar model consists of 24 elements. 
The individual contributions 
to the Rosseland opacity along $ 0\leq R \leq 0.8R_{\odot} $, due to individual
elements in the mixture and due to all metallic elements combined, are given in \autoref{fig:elements_contri_kappa}. It is seen that the most contributing metallic elements are iron and oxygen, while 
other dominant elements (with a contribution larger than 10\% over some range in solar interior) are silicon, neon, magnesium and sulfur.
As mentioned in the introduction, it is evident that metals have a major opacity fraction throughout the solar interior, rising from about $ 40\% $, at the core to about  $ 90\% $ near the CZB.
The metallic opacity at the core is mainly due to photoexcitation and photionization of the iron K-shell, while near the CZB it is mainly due to photoexcitation and photionization of the oxygen, neon and magnesium K-shell and of the iron M and L shells (\cite{blancard2012solar,colgan2013light,colgan2016new,krief2016sun_sta_apj}).

In the STAR code, Voigt profiles are used to describe the spectral envelopes of the super-transition-arrays.
The Gaussian width
results from Doppler broadening and from a statistical broadening due to the huge number of lines in each array. This statistical width is due to the fluctuations in the occupation numbers of the various configurations in each superconfiguration, and due to the unresolved-transition-array (UTA) widths of the various configurations (\cite{moszkowski1962energy,bauche1979variance,bauche1985variance,bar1995effect,gilleron2008impact,gilleron2011corrections,pain2009effect,bauche2015atomic,krief2015variance}).
The Lorentzian width contains natural broadening and
electron impact (Stark) broadening via the widely used semi empirical formulas by
\cite{dimitrijevic1987simple}.

Opacity calculations by several groups (\cite{iglesias1995discrepancies,colgan2013light,colgan2016new,blancard2012solar,krief2016sun_sta_apj}), have indicated much smaller K-shell line widths than obtained by the OP (\cite{badnell2005updated}), specifically for oxygen and magnesium near the CZB ($ T=192.9eV $, $ n_{e}=10^{23}cm^{-3} $).
On the other hand, these opacity calculations also show that the OP opacity of iron in the spectral region of the M-shell is significantly lower due to smaller OP populations of ionic states with at least one electron in the M-shell,
for which ground and excited configurations must have a non-empty M-shell.
These configurations contribute significantly to the opacity in the M-shell spectral region
(see \cite{blancard2012solar,colgan2016new,krief2016sun_sta_apj}).
Such lower OP populations may result from the different ionization balance modeling implemented in STAR and OP, or simply due to a truncation of relevant excited configurations in the OP calculation.
We note that, as was shown in detail by \cite{krief2015effect}, except for heavy elements such as iron and nickel nearby the CZB, there is a good agreement between STAR and OP ionic state distributions.

\begin{figure}[H]
	\centering
	\resizebox{\textwidth}{!}{
		\begin{minipage}{\textwidth}
			\resizebox{0.24\textwidth}{!}{\includegraphics{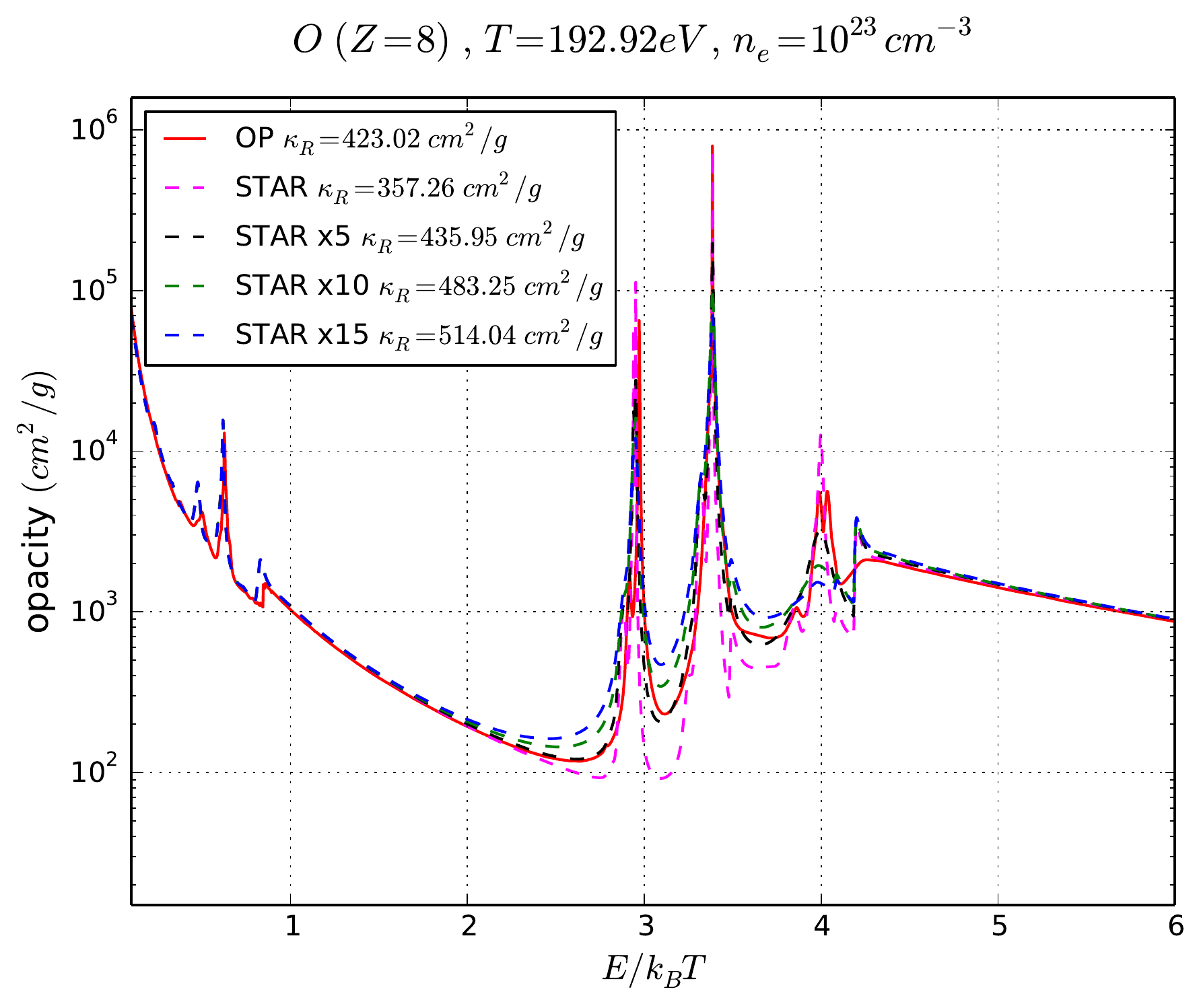}}
			\resizebox{0.24\textwidth}{!}{\includegraphics{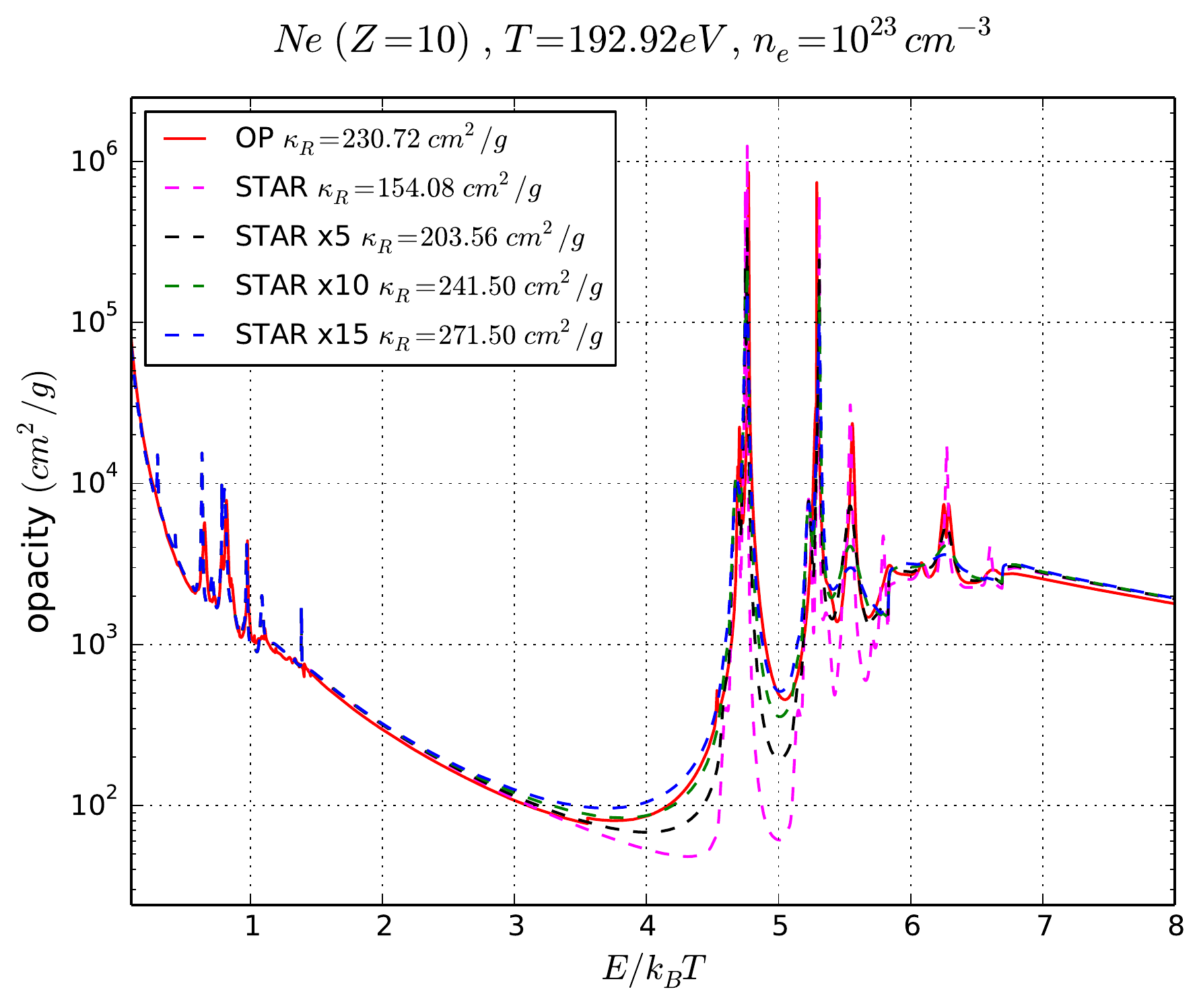}}
			\resizebox{0.24\textwidth}{!}{\includegraphics{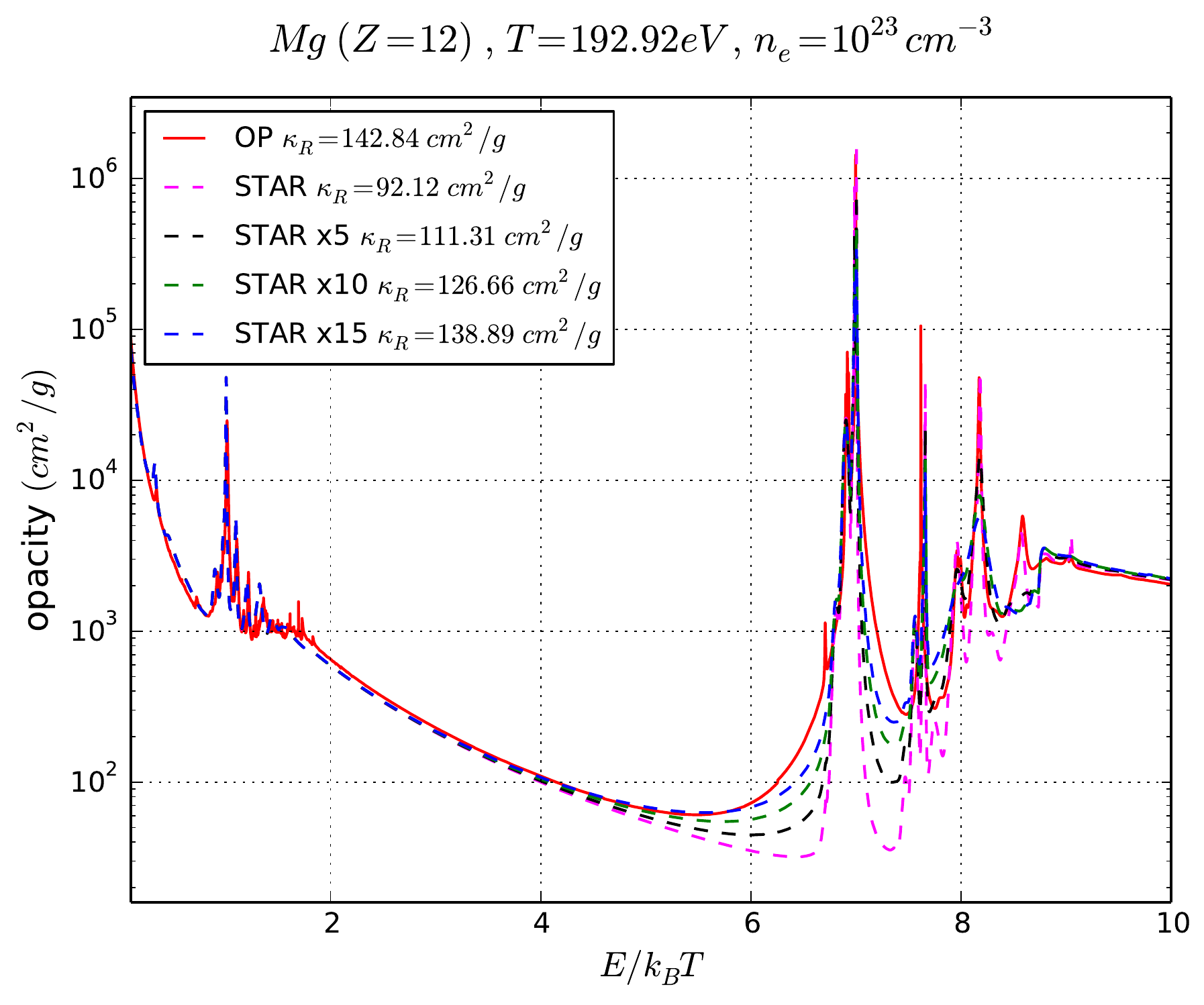}}
			\resizebox{0.24\textwidth}{!}{\includegraphics{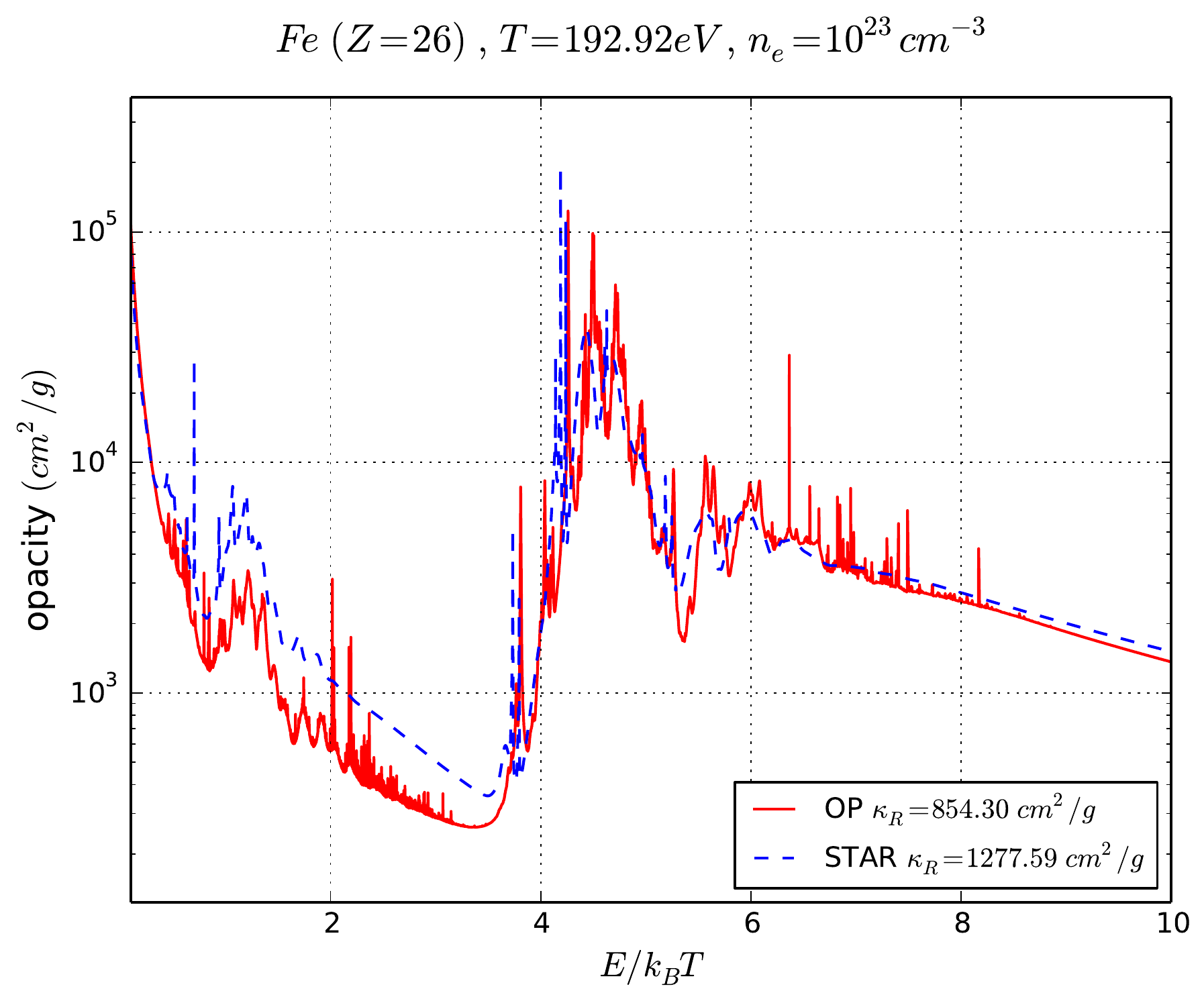}}
			\newline
			\resizebox{0.24\textwidth}{!}{\includegraphics{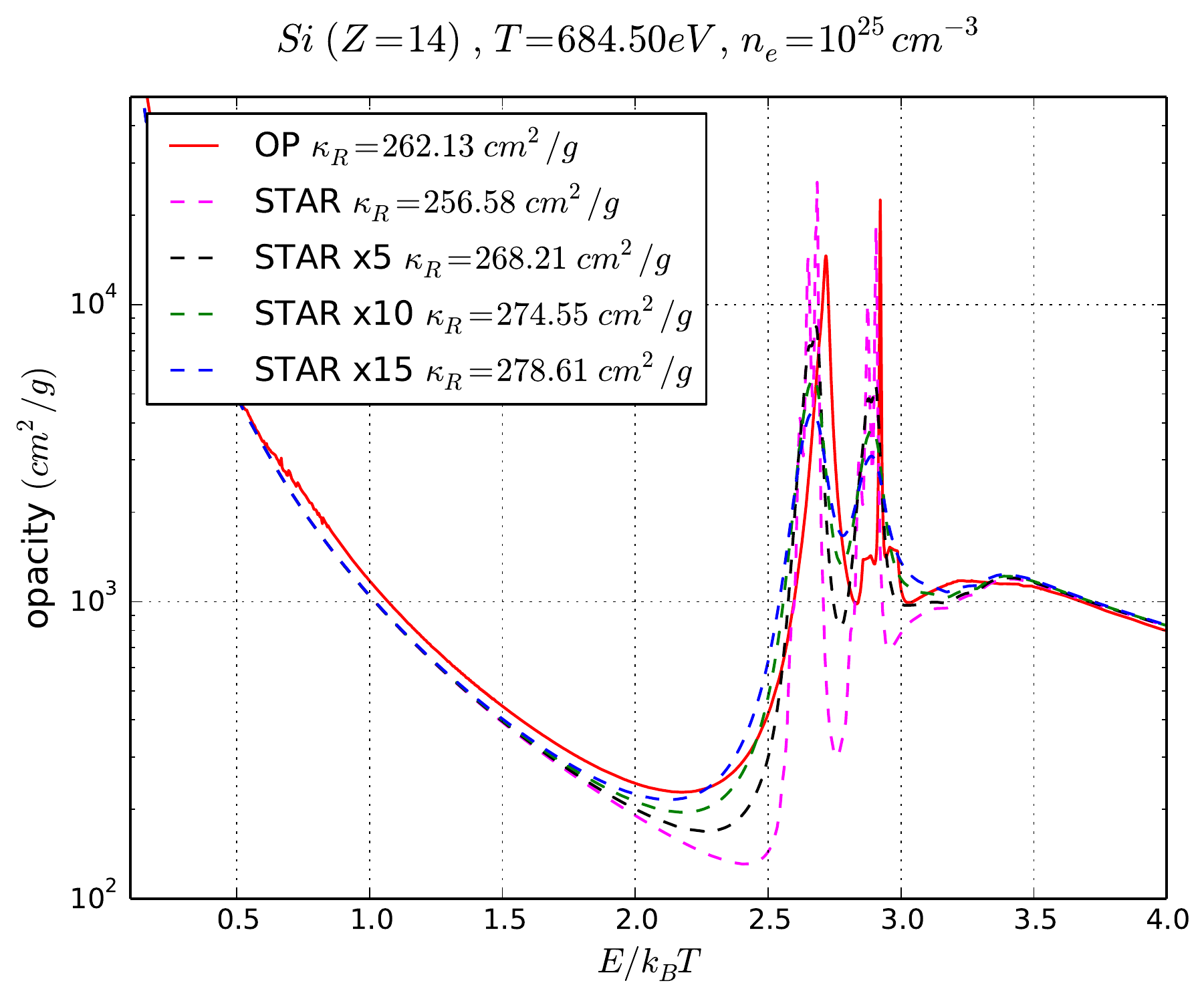}}
			\resizebox{0.24\textwidth}{!}{\includegraphics{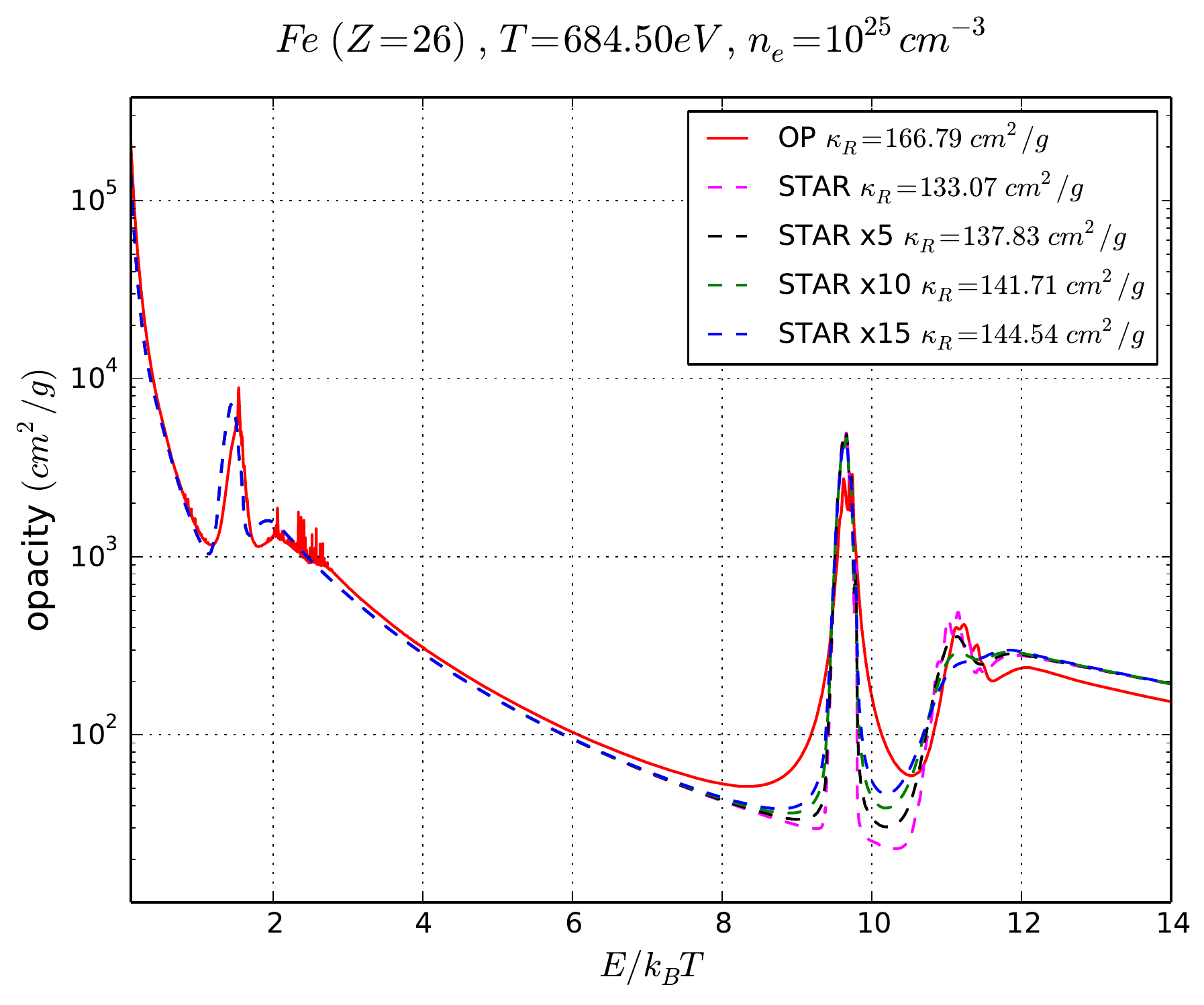}}			
			\resizebox{0.24\textwidth}{!}{\includegraphics{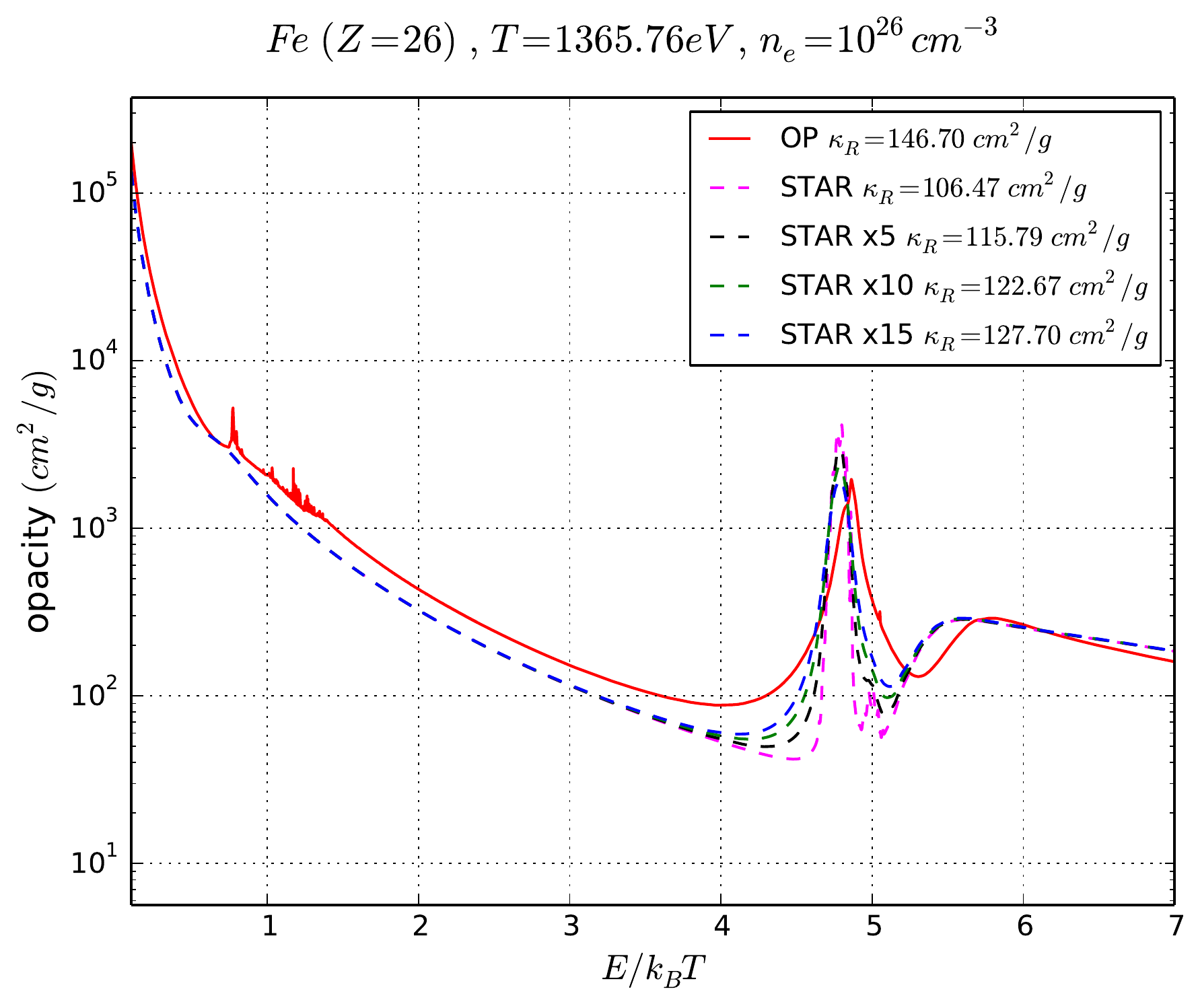}}
			\resizebox{0.24\textwidth}{!}{\includegraphics{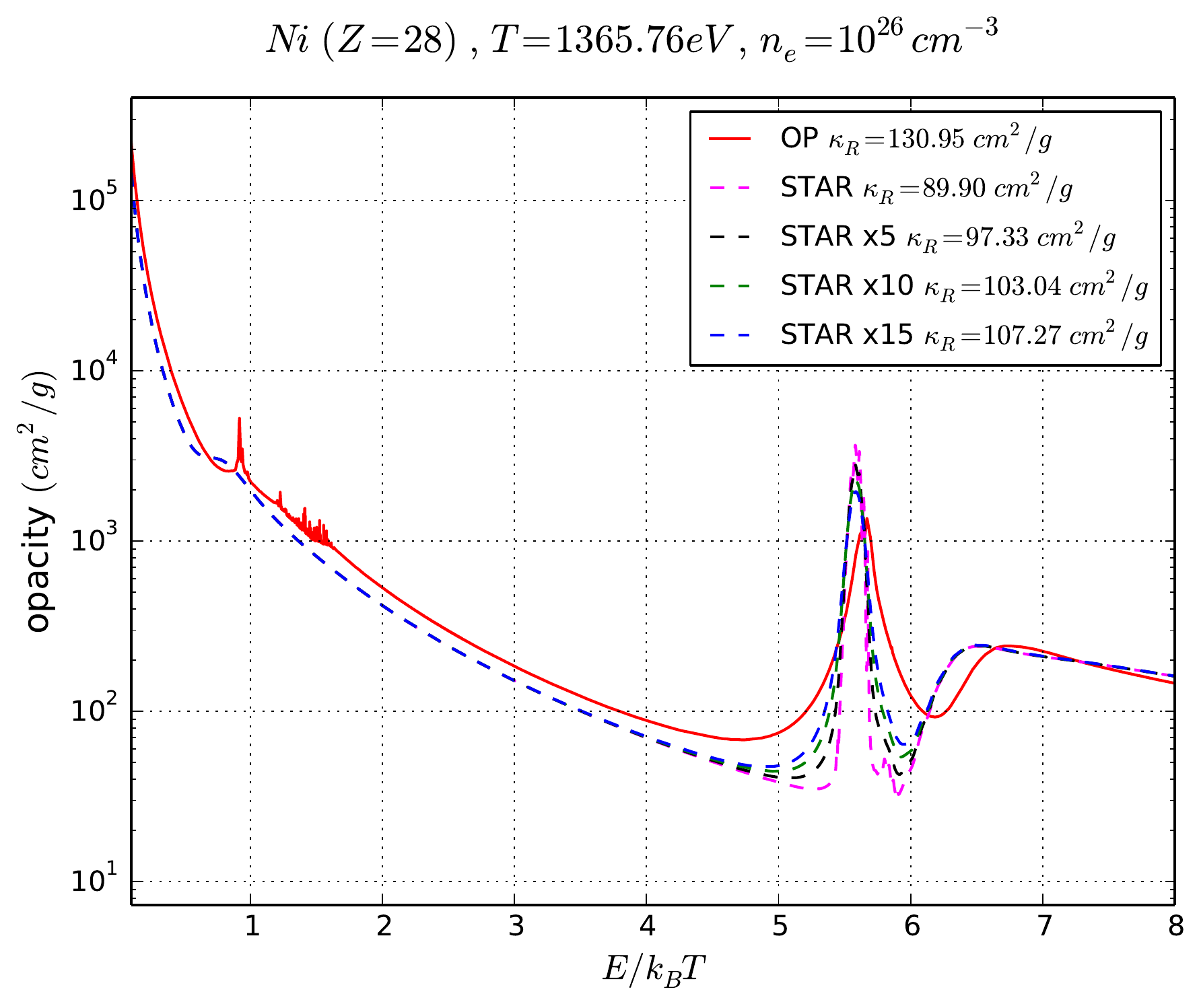}}			
			\caption{
				OP and STAR opacity spectra for oxygen, neon, magnesium and iron at $ T=192.9eV $, $ n_{e}=10^{23}cm^{-3} $ (found nearby the CZB); for silicon and iron at $ T=684.5eV $, $ n_{e}=10^{24}cm^{-3} $ (found at $ R\approx0.25 R_{\odot}$) and for iron and nickel at $ T=1365.76eV $, $ n_{e}=10^{26}cm^{-3} $ (found nearby the solar core).
				STAR spectra with the K-shell Stark line-widths multiplied by a factor of $ 5,10 $ and $ 15 $ are also shown (except for iron at $ T=192.9eV $, $ n_{e}=10^{23}cm^{-3} $ for which the K-shell lines do not exist in the spectral range of interest).
				}
			\label{fig:spec_braods_op_sta15}
		\end{minipage}
	}
\end{figure}

These differences have a major effect on the Rosseland opacities of the individual elements, but a rather small effect on the total solar mixture opacity.
In \autoref{fig:spec_braods_op_sta15} STAR and OP spectral opacities are given for oxygen, neon, magnesium and iron at $ T=192.9eV $, $ n_{e}=10^{23}cm^{-3} $ (nearby the CZB); for silicon and iron at $ T=684.5eV $, $ n_{e}=10^{24}cm^{-3} $ (found at $ R\approx0.25 R_{\odot}$) and for iron and nickel at $ T=1365.76eV $, $ n_{e}=10^{26}cm^{-3} $ (nearby the solar core).
STAR spectra with the K-shell Stark line-widths multiplied by arbitrary factors of $ 5,10 $ and $ 15 $ are also shown.
The differences described above between the spectra at the CZB, are evident from the figure.
It appears that the discrepancy in the K-shell widths increases with the atomic number. For example, nearby the CZB, the discrepancy  is about a factor of 5 for oxygen, about a factor of 10 for neon and a factor larger than 15 for magnesium. 
Large discrepancies in the K-shell widths appear also for silicon and iron at $ R\approx0.25 R_{\odot}$ and for iron and nickel nearby the solar core, for which there is also a large discrepancy in M-shell photoionization spectra. 
On the other hand, for iron nearby the CZB, the K-shell lines do not contribute to the opacity and there is a relatively good agreement in the L and M-shell line widths, although the M-Shell opacity of STAR is much higher, due to the higher populations of the relevant configurations, as explained above.
In conclusion, there is a large discrepancy in the K-shell line widths between OP and STAR (as well as other atomic codes), 
which is mainly due to differences in the Stark widths.

\begin{figure}[H]
	\centering
	\resizebox{\textwidth}{!}{
		\begin{minipage}{\textwidth}
			\subfloat[Stark widths$ \times2 $]{
				\includegraphics[width=60mm]{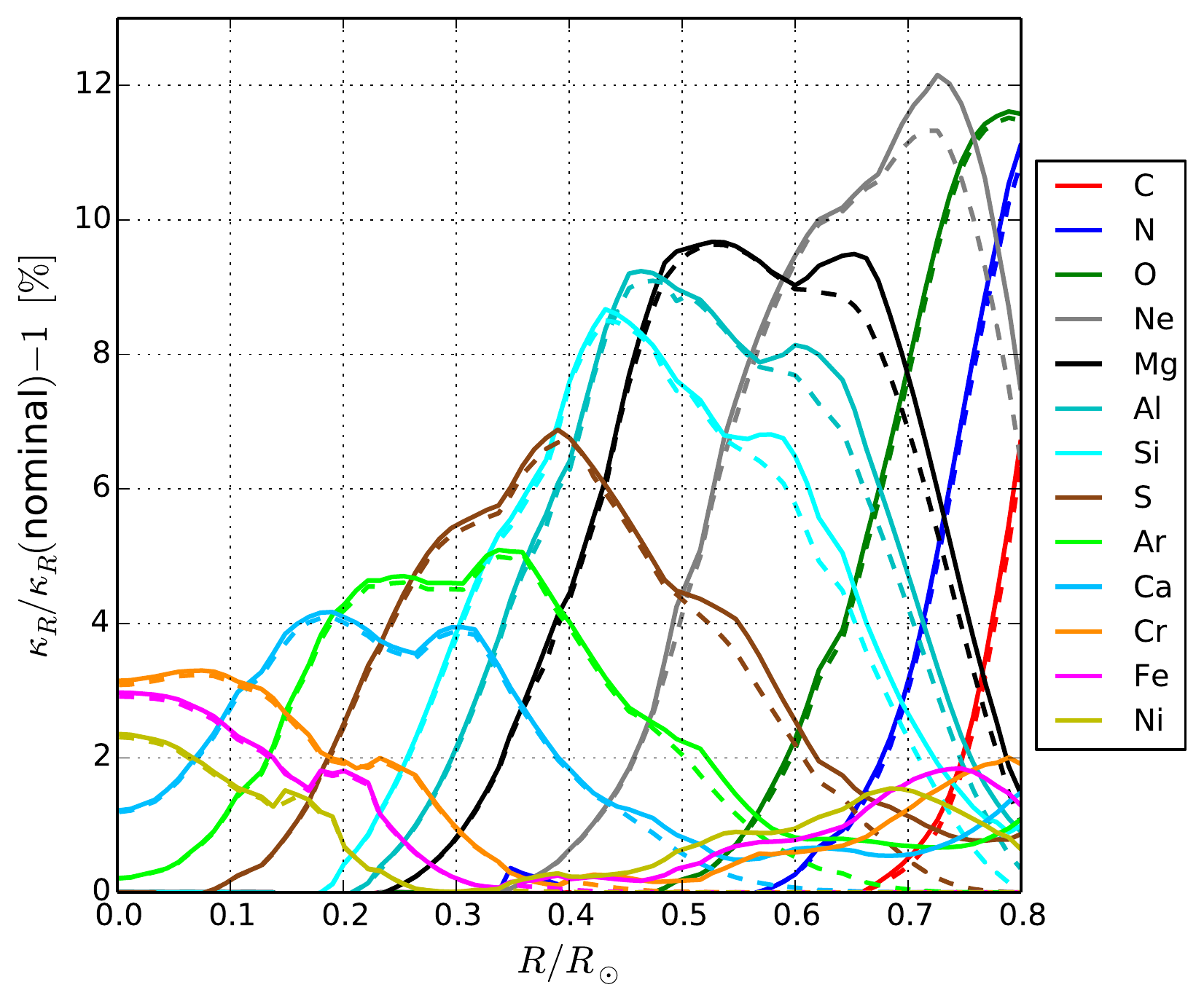}
			}
			\subfloat[Stark widths$ \times5 $]{
				\includegraphics[width=60mm]{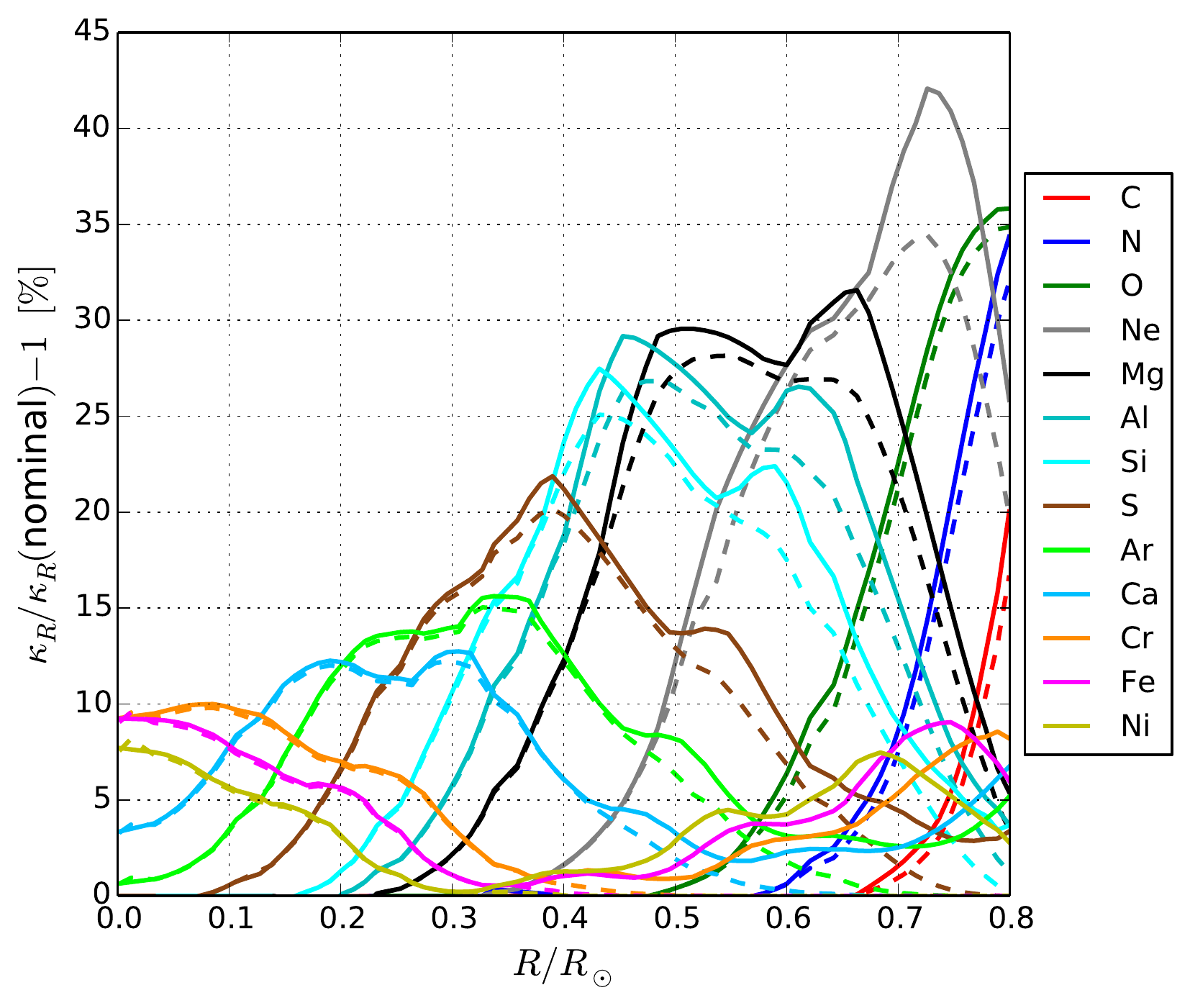}
			}
			\subfloat[Stark widths$ \times10 $]{
				\includegraphics[width=60mm]{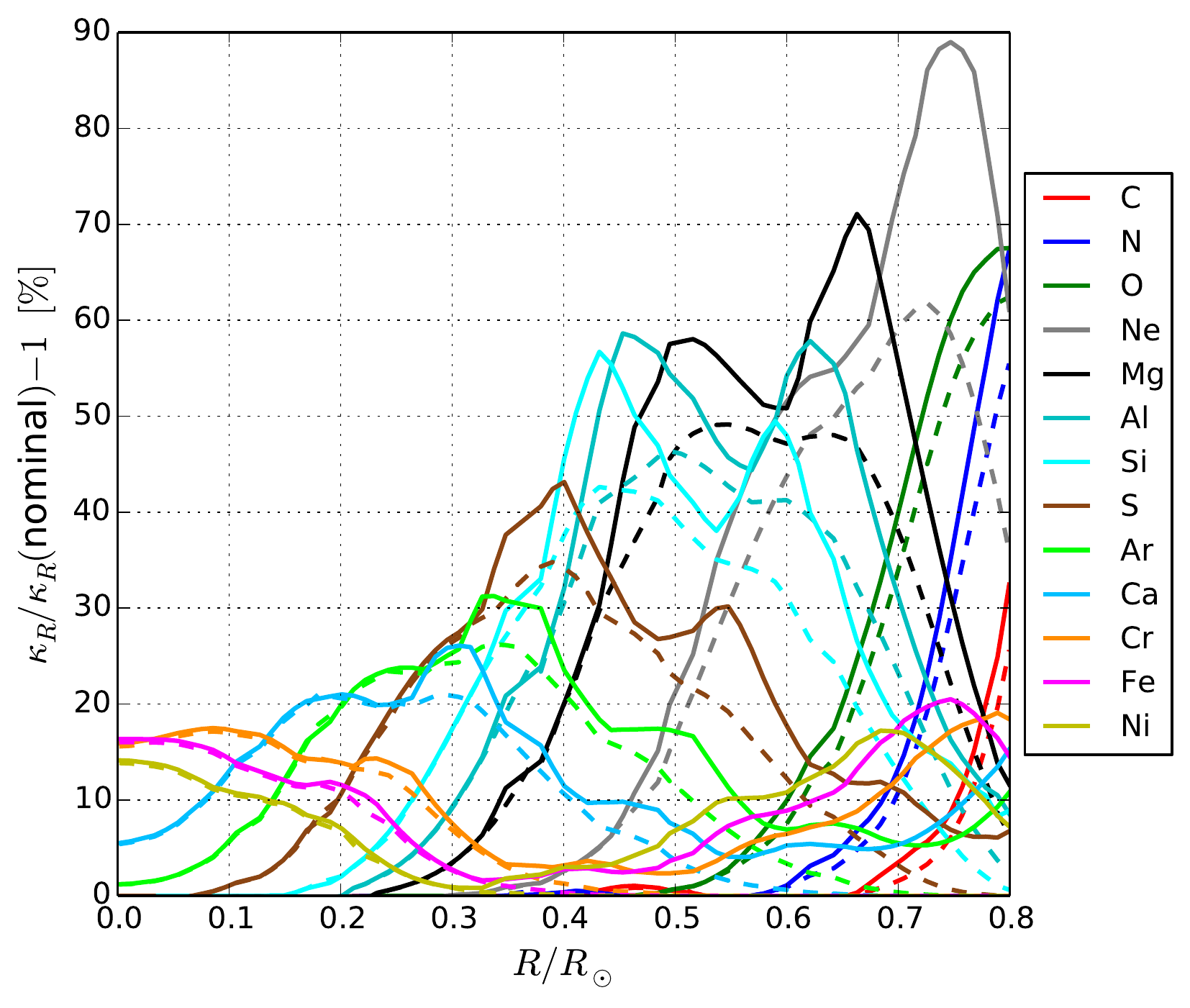}
			}
			\newline
			\subfloat[Stark widths$ \times15 $]{
				\includegraphics[width=60mm]{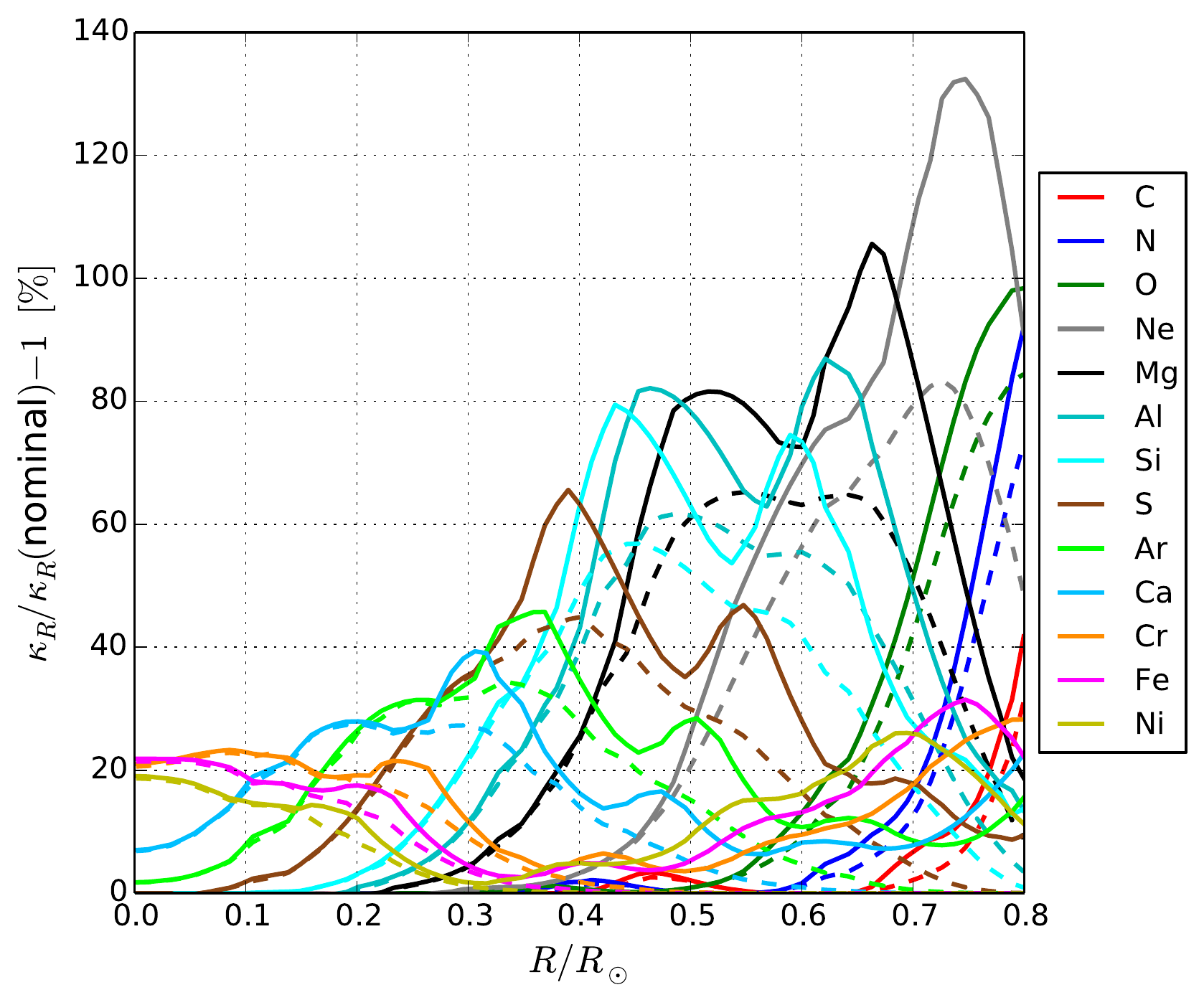}
			}
			\subfloat[No Lorentzian widths]{
				\includegraphics[width=60mm]{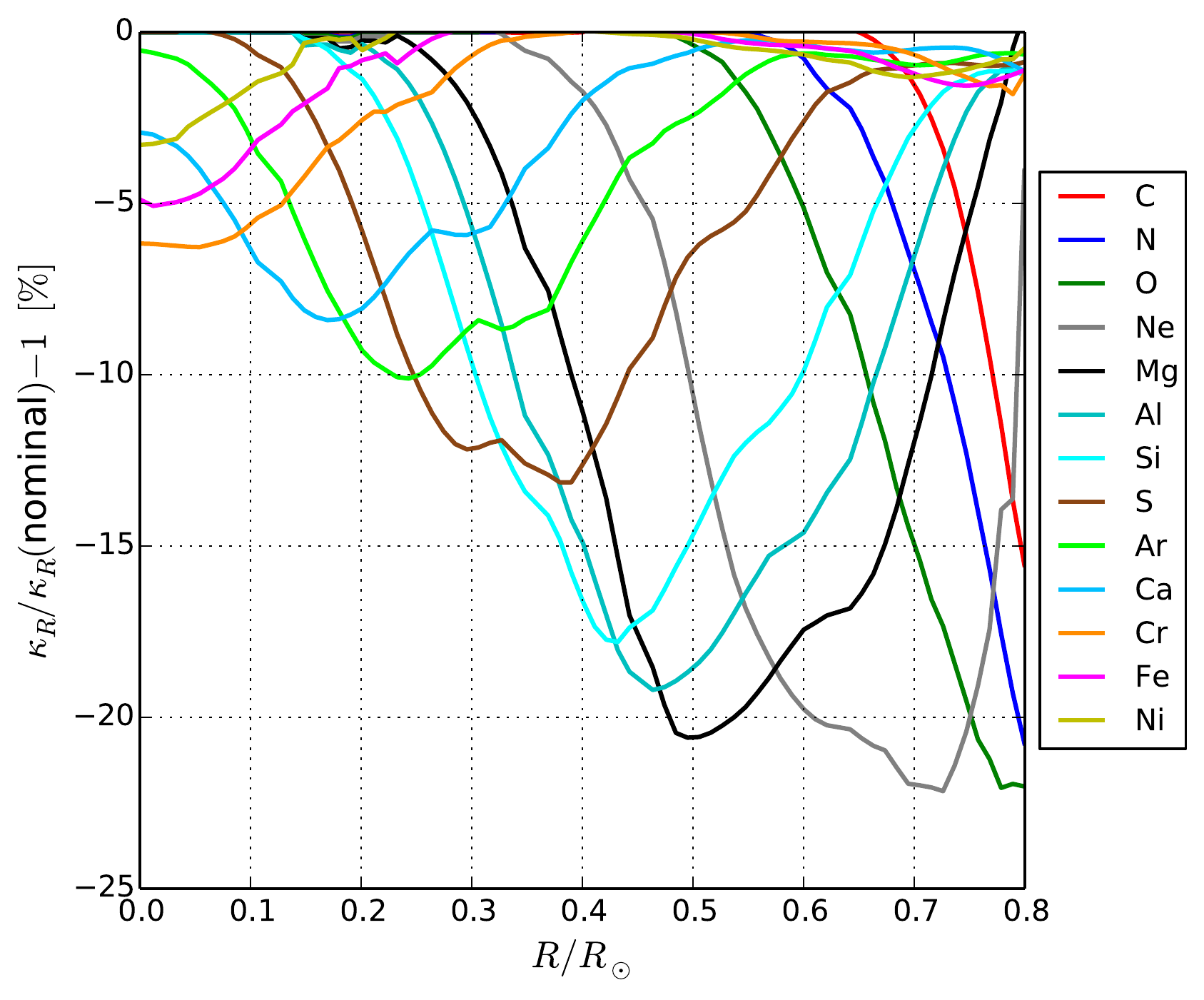}
			}
			\subfloat[No UTA widths]{
				\includegraphics[width=60mm]{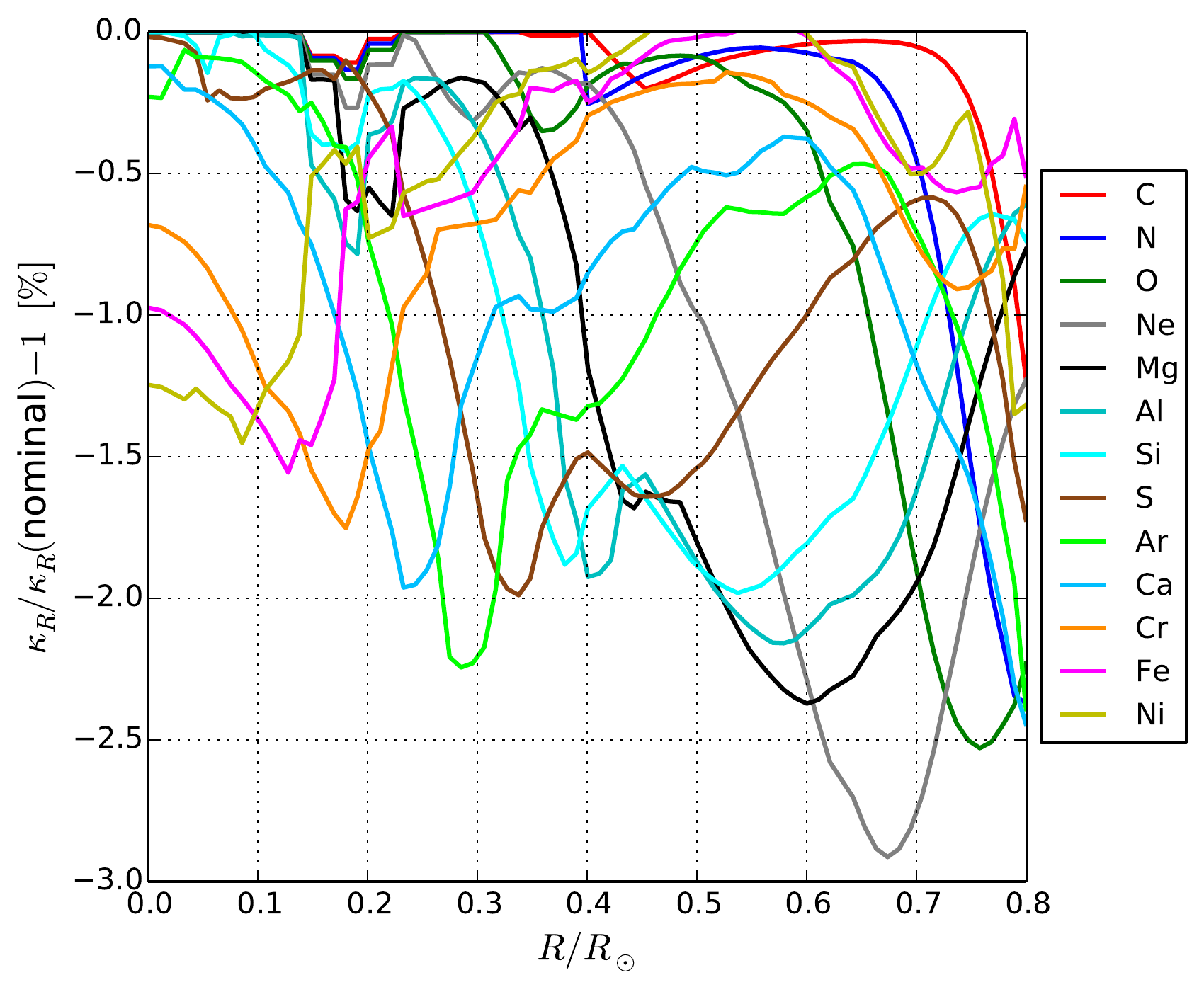}
			}
			\newline
			\caption{
				Relative differences in the opacity of various elements,
				between STAR calculations with altered line widths, and the nominal STAR calculation.
				In sub-figures (a)-(d), the Stark line widths were multiplied by $ 2,5,10 $ and $ 15 $, respectively,  for all lines (solid curves) and for K-shell lines only (dashed curves).
				Calculations with zero Lorentzian widths, and without UTA widths are
				given in sub-figures (e),(f), respectively.
				The calculations are made for $ 0\leq R\leq 0.8R_{\odot} $ across the solar interior.
			}
			\label{fig:sens_dk_elements}
		\end{minipage}
	}
\end{figure}

In order to estimate the sensitivity of the opacity of elements in the solar mixture to line broadening, we have performed  solar opacity calculations with the Stark widths of all lines multiplied by factors arbitrary of $ 2,5,10 $ and $ 15 $. The calculation was also performed with a multiplication of K-shell lines widths only. In addition, we have performed a calculation
 with zero Lorentzian widths (that is, with Gaussian line shapes) and a calculation without the UTA widths, which is the part of the total Gaussian width that is due to the individual widths of the various UTAs in each STA. The relative differences between these calculations and the nominal STAR calculation are shown in \autoref{fig:sens_dk_elements}.
It is seen that indeed, the Lorentzian line broadening has a major effect on the opacity of the individual elements.
We note that maxima which appear in the sensitivities are due to  spectral lines and photoionization edges that cross the Rosseland peak (located at $ \approx 3.8k_{B}T $), as the temperature changes throughout the solar interior. This also explains why these maxima occur at a larger radii for lower $ Z $ elements: since the line energies are lower for lighter elements, the contribution of the lines is higher at lower temperatures - which occur at larger radii. For Stark widths multiplied by $ 2 $ and $ 5 $ (Figs. \ref{fig:sens_dk_elements}a-\ref{fig:sens_dk_elements}b), it is seen that the opacity changes mainly due to the broadening of the K-shell lines, for the lighter elements with $ Z\lesssim 15 $. For the heavier elements, in the range $ R\lesssim0.25 R_{\odot} $ the opacity  changes
mainly due to broadening of the K-shell lines, while for
$ R\gtrsim  0.5 R_{\odot}$
the L and M-shell line widths affect the opacity. 
For larger multiplication of the Stark widths (by factors of $ 10$ and $15 $, shown in Figs. \ref{fig:sens_dk_elements}c-\ref{fig:sens_dk_elements}d), there is a significant contribution to the opacity due to non K-shell lines, for elements with $ Z\lesssim 15 $, explained by the large wings of the Voigt profiles of such lines which increases the background opacity (which is due to inverse photoabsorption, bremsstrahlung and scattering), at the vicinity of the Rosseland peak.
It is seen from \autoref{fig:sens_dk_elements}e that the nominal Lorentz widths also have a significant effect on the elemental opacities.
Finally, \autoref{fig:sens_dk_elements}f shows that
 the UTA widths have a small effect ($ \lesssim  3\%$), which means that the Gaussian widths are dominated by the variance of the transition energies of the UTAs within super-transition arrays, and not by the individual widths of the UTAs.
We note that in general, the heavier elements are less sensitive to  line broadening, since their spectra is composed of a larger number of overlapping lines, so that the widths of  individual lines  are less important than the statistical widths due to the different energies of the various overlapping lines.

\begin{figure}[h]
	\centering
	\resizebox{0.5\textwidth}{!}{\includegraphics{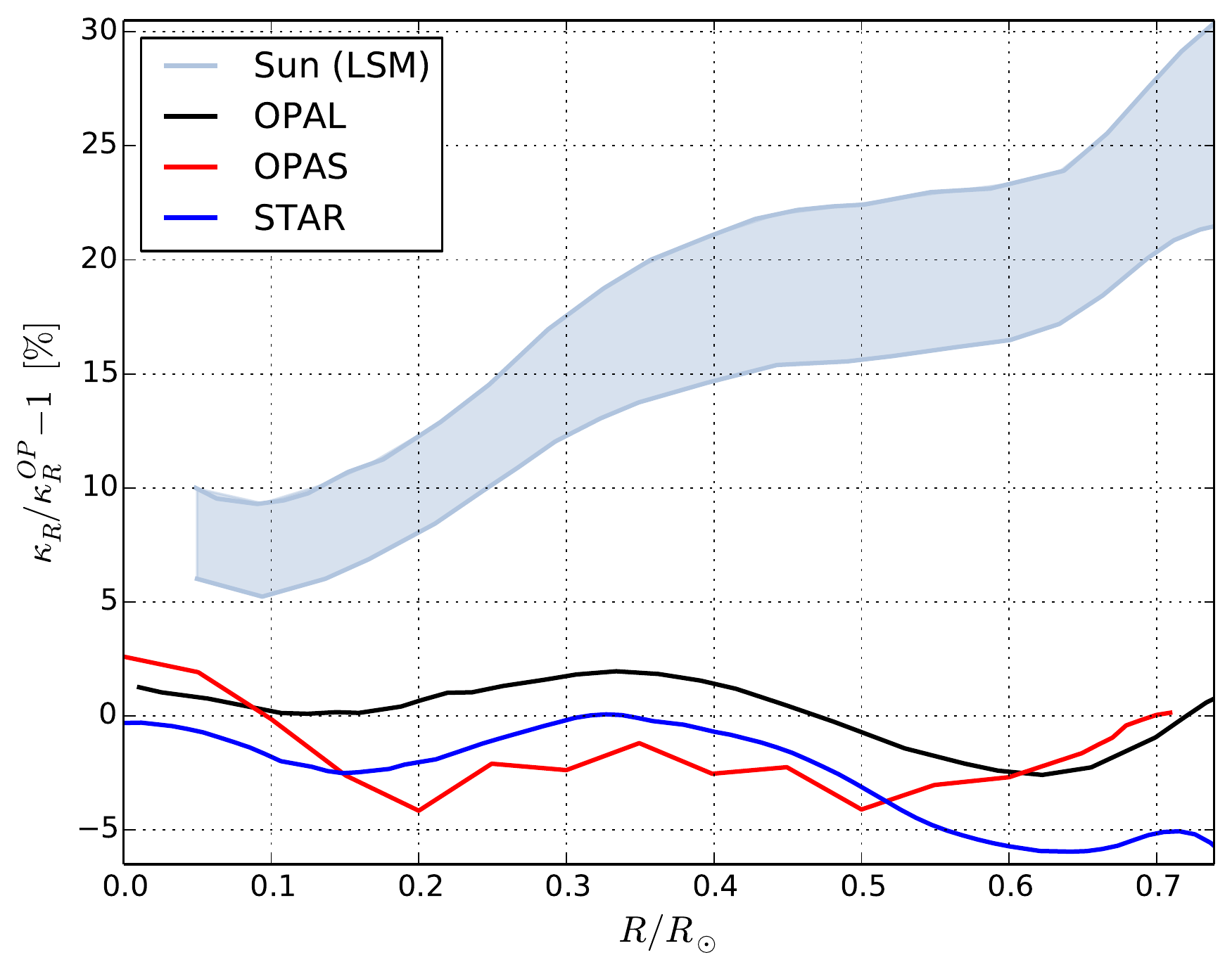}}
	\caption{
			Relative differences $ \kappa_{R}/\kappa^{OP}_{R}-1 $, between OP 
		and OPAL (in black, data is taken from \cite{villante2014chemical}), OPAS (in red, data is taken from \cite{blancard2012solar}) and STAR (in blue),
		across the solar radiative zone.
		The shaded area represents the range of the variation in the opacity profile, required for helioseismic and neutrino observations to agree with predictions, calculated by a linear solar model (data is taken from \cite{villante2014chemical}). 
		}
	\label{fig:dk_sensitivity_op}
\end{figure}

\begin{figure}[h]
	\centering
	\resizebox{0.5\textwidth}{!}{\includegraphics{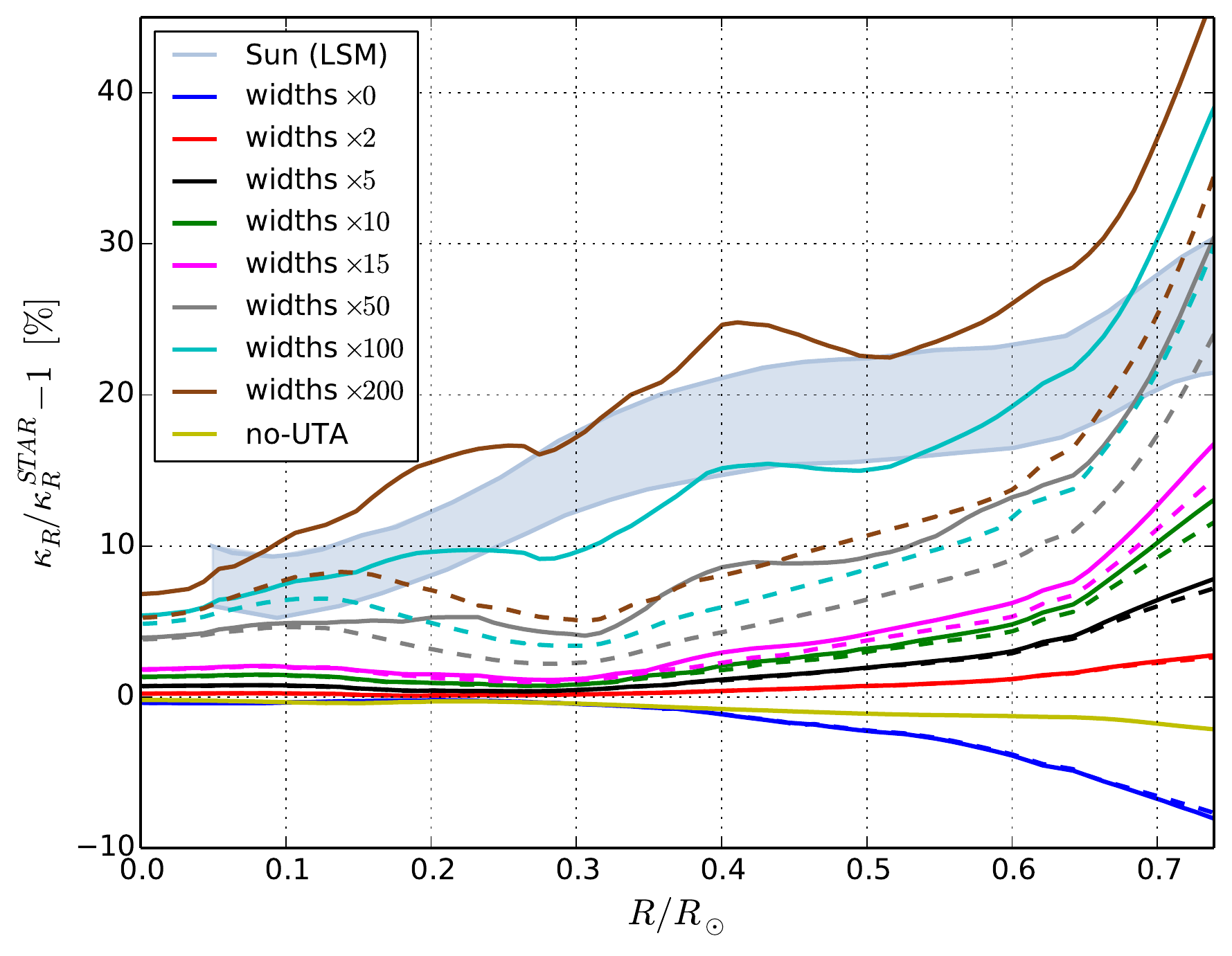}}
	\caption{
			Relative differences $ \kappa_{R}/\kappa^{STAR}_{R}-1 $, between a nominal STAR calculation and STAR calculations with altered line-widths. Calculations are performed with no Lorentzian widths (in blue), with Stark widths of all lines (solid curves) and only K-shell lines (dashed curves) multiplied by $ 2,5,10, 15, 50,100 $ and $ 200 $ (in red, black, green, magenta, grey, turquoise and brown, respectively),  and with zero UTA widths (in brown). 
		}
	\label{fig:dk_sensitivity_star}
\end{figure}

\cite{villante2014chemical} have calculated
the range of the variation in the opacity profile, required for helioseismic and neutrino observations to agree with standard solar models, calibrated with the recent AGSS09 abundances. 
The range of the required opacity variation is shown in \autoref{fig:dk_sensitivity_op}, in the form of a band.
It is important to note that in order to reconcile the observed and predicted sound speed profile and convection boundary radius,
variations in the opacity must only have a shape lying within the variation band. The magnitude of the opacity variation is required to reconcile the observed neutrino fluxes and surface helium abundance.
 The relative differences between OP to OPAL, OPAS and STAR, which are much smaller than the required variation, are also shown in \autoref{fig:dk_sensitivity_op}.
Relative differences between a nominal STAR calculation and STAR calculations with altered line widths are shown in \autoref{fig:dk_sensitivity_star}.
It is evident that the sensitivity to the values of the line widths is larger nearby the CZB than near the central regions.
It is also seen that the curves representing width multiplications of $ 2,5,10 $ and $ 15 $, result mainly due to the K-shell lines. This is understood from the fact that nearby the CZB, abundant metals (oxygen, neon, magnesium) have their K-shell lines nearby the Rosseland peak, while heavier elements such as iron have their L-shell lines, which are less sensitive to line-broadening, at the vicinity of the Rosseland peak. On the other hand, in the central regions, due to the high density, only few heavy metals (such as iron and nickel) have their K-shell not pressure ionized and nearby the Rosseland peak. 
For the larger width multiplications of $ 50,100 $ and $ 200 $, a substantial part of the opacity variation is due to non K-shell lines, which, due to their very large widths, have spectral contributions near the Rosseland peak. As was analyzed in detailed by \cite{krief2016sun_sta_apj},
the maxima and minima in the opacity variations in \autoref{fig:dk_sensitivity_star}
result from lines of various elements that cross the Rosseland peak as the temperature varies along the solar thermodynamic path.
Finally we note that, as expected, it is seen from \autoref{fig:dk_sensitivity_star} that the UTA widths have a very small effect ($ \lesssim 2\% $) throughout the solar radiative zone.

\section{Discussion and Summary}
Different calculations of solar opacities by various groups (i.e. OP, OPAL, OPAS, ATOMIC and STAR) show a discrepancy of $ \lesssim 5 \% $, which is much smaller than the  variations required to bring back the agreement between
helioseismology and SSMs calibrated with the updated low-Z solar composition.
However, an element
by element comparison with OP shows large
differences  ($ \approx \pm 40 \% $)  nearby the CZB.
On the other hand, comparisons
between atomic codes other than OP show
a good agreement ($ \approx $few$\% $), which indicates that these atomic codes use similar approximations for the influence of the plasma environment on the emitting ions (such as line broadening), which may differ than the approximations implemented by OP.
It must also be stressed that the small \textit{discrepancy} in the solar mixture opacity between different atomic calculations is by no means a measure of the \textit{accuracy} of these calculations, since similar physical models and approximations are used.

Opacity calculations at stellar interior conditions have never been validated experimentally.
A recent work on experimental opacities indicates
that theoretical opacity calculations, which are the only available
source of radiative opacities in solar models,
might in fact be systematically wrong.
Opacity spectra of iron have been measured by \cite{bailey2015higher}, for the first time, at conditions
that closely resemble those at the CZB.
The measured Rosseland mean is larger by $ \approx  60 \% $ than all predicted values by available atomic calculations.
No satisfactory explanation to this discrepancy is yet known.

In this work, line-broadening, which results from the extremely complex interaction of the plasma environment with the emitting ions, 
was pointed out as a source of uncertainty in the calculation of solar opacities. First, as was shown in several recent works (\cite{blancard2012solar,colgan2016new,krief2016sun_sta_apj}), it was stressed that K-shell line widths given by OP are much larger than obtained from other calculations of spectral opacities of metals within the solar interior.
The differences in the widths, which were shown to increase with the atomic number, range from a factor of $ \approx 5 $ to a factor of $ \approx 15 $.
Rosseland opacities were calculated by STAR for a solar model calibrated with the recent AGSS09 photospheric abundances and the
sensitivity of the elemental opacities of various elements to line-broadening was studied throughout the solar interior.
Such moderate width multiplications (by factors of $ 2,5,10$ and $ 15 $)
were shown to have a non-negligible effect on the solar opacity profile, which is shown to result mainly from K-shell lines.
The resulting opacity variations are
found to mimic the behavior
of the required opacity variation profile of the present day sun, as imposed by helioseismic and neutrino observations via the framework of the linear solar model. 
Although the missing opacity nearby the CZB is somewhat reduced
 due to such uncertainty in line broadening, it is still large in the internal regions where the line broadening effect is less than $ 5\% $ (see \autoref{fig:dk_sensitivity_star}).

	Of course, the actual uncertainty in the line widths cannot be represented by the differences between atomic codes that use very approximate line broadening schemes, which are experimentally untested at solar interior conditions.
In order to see just how much the Stark broadening would need to be increased in order to actually get into the required range of opacity variation, 
we have also performed calculations with larger width multiplications of $ 50,100 $ and $ 200 $. 
As seen from \autoref{fig:dk_sensitivity_star}, a width multiplication of a value between $ 100 $ and $ 200 $ for all lines may recover the missing opacity in both shape and magnitude, in the region $ R\lesssim 0.6 R_{\odot} $.
However, at larger radii and specifically, nearby the CZB, a lower multiplication of about $ 50 $ is required. Hence, we conclude that the line multiplication that fully recovers the missing opacity must be non-uniform throughout the solar interior.
We believe that the largest multiplication factor that can reasonably be entertained in \autoref{fig:dk_sensitivity_star} is quantitatively unknown.
In fact, the unknown uncertainty due to line broadening may vary independently and in a very complicated way for different transition lines, atomic numbers, chemical mixtures, temperatures and densities. Moreover, in this work we have only explored the uncertainty in the \textit{width} of a given approximate form of the \textit{line shapes}, without addressing the more sophisticated task of exploring the uncertainty and sensitivity in the choice of line shapes.

This signifies the need for a better theoretical characterization of the uncertainties in the modeling
	  and implementation
	  of the
	   line broadening phenomena by state of the art atomic codes.	
If the uncertainty in the line widths at solar interior conditions is about a factor of $ 10 $ or less, then due to the higher densities prevailing in the inner regions, other sources of uncertainty in opacity calculations are in order.
These may well be related to the different modeling of the plasma environment  surrounding the emitting ions, which may give rise to large differences in the ionic states distribution (\cite{iglesias1995discrepancies,blancard2012solar,colgan2016new,krief2016sun_sta_apj}), the different boundary conditions in the framework of ion-sphere models, and corrections such as short range ion-ion and electron-ion correlations 
(\cite{rozsnyai1991photoabsorption,starrett2012average,murillo2013partial,Chihara201638}).
We also note that the differences between the recently measured
(\cite{bailey2015higher,nagayama2016calibrated})
 and calculated iron spectra seem to be more complicated than a trivial difference in the line widths.

The authors thank Aldo Serenelli and Francesco Villante for providing solar data and for
useful suggestions and comments.
\acknowledgments{

}
	
\bibliographystyle{apj}

\newpage

\end{document}